\documentclass[preprint, nobibnotes, amsmath,amssymb, aps, superscriptaddress, longbibliography]{revtex4}
\usepackage{graphicx}
\usepackage{dcolumn}
\usepackage{bm}
\usepackage[export]{adjustbox}
\usepackage{xcolor}
\usepackage{booktabs}
\usepackage{multirow}
\usepackage[english]{babel}
\usepackage{physics}
\usepackage[ruled,vlined]{algorithm2e}
\usepackage{float}
\usepackage{times}
\usepackage[pdftex, pdftitle={Article}, pdfauthor={Author}]{hyperref}
\usepackage{hyperref}
\begin{document}
\title{Deep learning-based variational autoencoder for classification of quantum and classical states of light}
\author{Mahesh Bhupati$^{\dagger}$}
\affiliation{Laboratory of Optics of Quantum Materials (LOQM), Department of Physics, Indian Institute of Technology Bombay, Powai, Mumbai 400076, India}
\author{Abhishek Mall$^{\dagger}$}
\affiliation{Max Planck Institute for the Structure and Dynamics of Matter, 22761 Hamburg, Germany}
\affiliation{Center for Free Electron Laser Science, 22761 Hamburg, Germany}
\author{Anshuman Kumar}
\affiliation{Laboratory of Optics of Quantum Materials (LOQM), Department of Physics, Indian Institute of Technology Bombay, Powai, Mumbai 400076, India}
\affiliation{Centre of Excellence in Quantum Information, Computation, Science and Technology, Indian Institute of Technology Bombay, Powai, Mumbai 400076, India}
\author{Pankaj K. Jha\footnote{Correspondence and requests for materials should be addressed to P.K.J (pkjha@syr.edu)}} 
\affiliation{Quantum Technology Laboratory $\langle Q|T|L\rangle$, Department of Electrical Engineering and Computer Science, Syracuse University, Syracuse, NY, USA}
\begin{abstract}
Advancements in optical quantum technologies have been enabled by the generation, manipulation, and characterization of light, with identification based on its photon statistics. However, characterizing light and its sources through single photon measurements often requires efficient detectors and longer measurement times to obtain high-quality photon statistics. Here we introduce a deep learning-based variational autoencoder (VAE) method for classifying single photon added coherent state (SPACS), single photon added thermal state (SPACS), mixed states between coherent/SPACS and thermal/SPATS of light. Our semi-supervised learning-based VAE efficiently maps the photon statistics features of light to a lower dimension, enabling quasi-instantaneous classification with low average photon counts. The proposed VAE method is robust and maintains classification accuracy in the presence of losses inherent in an experiment, such as finite collection efficiency, non-unity quantum efficiency, finite number of detectors, etc. Additionally, leveraging the transfer learning capabilities of VAE enables successful classification of data of any quality using a \textit{single} trained model. We envision that such a deep learning methodology will enable better classification of quantum light and light sources even in the presence of poor detection quality.
\end{abstract}
\maketitle

\section{\label{sec:intro}Introduction}

\noindent The development of novel technologies that allow for the generation, manipulation, and characterization of quantum states of light has led to a substantial advancement in the field of quantum optics in recent years~\cite{monroe2013scaling,kimble2008quantum,kwon2019nonclassicality,van2006hybrid, flamini2018photonic, moreau2019imaging,leibfried2004toward} with applications spanning quantum communication, cryptography, and sensing~\cite{dowling2008quantum,degen2017quantum,kremer1995quantum}. Different quantum states of light have distinctive properties that make them more suitable for particular applications~\cite{nielsen2002quantum,gisin2007quantum,ono2013entanglement,lemos2014quantum}. For instance, photon-added coherent states~\cite{agarwal1991nonclassical, agarwal1992nonclassical} denoted by $|\alpha, m\rangle$, where $m$ photons are added to the same mode as a coherent state $|\alpha\rangle$ via stimulated emission, have applications ranging from quantum sensing~\cite{schnabel2017squeezed, braun2014precision} and quantum key distribution~\cite{van2006quantum,barnett2018statistics}  to understanding fundamentals of quantum physics~\cite{parigi2007probing,tsakmakidis2018quantum,najafian2020identification}. Quasi-instantaneous identification and characterization of quantum states of light at the single photon level is one of the challenges of working with them. 

Recent research has demonstrated the potential of deep learning algorithms in nanophotonics~\cite{mall2020fast, mall2020cyclical,wiecha2021deep,piccinotti2020artificial,qu2023deep,zandehshahvar2022manifold}, quantum optics~\cite{havlivcek2019supervised} and quantum computing~\cite{garg2020advances}. The deep learning methods, dense neural networks (NNs), can be used to classify different quantum states of light accurately~\cite{you2020identification,ahmed2021classification,kudyshev2020rapid,gebhart2021identifying,gebhart2020neural}. These algorithms are trained on simulated or experimental data of varying average photon statistics and sample sizes allowing them to learn the features that distinguish one quantum light state from another. In scenarios where the photon count is low or the photon counting detector is inefficient, the classification of distinct quantum light sources can pose a challenge for deep learning algorithms due to a lack of information-rich statistics. To overcome this limitation, artificial neural networks have been utilized to identify quantum light sources, as demonstrated in prior work~\cite{gebhart2020neural,you2020identification}. As the average photon count and measurement binsize change, the information encoding also changes, making it necessary to train multiple models for different photon statistics scenarios. A successful distinction of quantum light sources requires exploring the statistical information change with varying average photon counts and binsize. To achieve this, a latent space representation that encapsulates information over multiple datasets could provide better information features for the classification of quantum light sources.

In this paper, we propose and implement a deep learning-based variational autoencoder (VAE) method to classify non-classical states of light generated by single photon addition to an initial coherent and thermal via stimulated emission. These non-classical states of light are also known as single photon-added coherent states~(SPACS) and single photon-added thermal states~(SPATS), respectively. Such sources are useful for high-bit-rate communication and quantum state engineering~\cite{lee2010quantum,wang2012photon,li2018generation,burch2015quantum,parigi2007probing,bellini2022demonstrating}. Our proposed method employs a semi-supervised learning approach to efficiently classify SPATS and SPACS. In this work, for simplicity, we focused on states of a single quantized field mode of coherent and thermal light. Using a variational autoencoder (VAE), the photon statistics features of the quantum states of light are mapped to a lower dimensional latent space, enabling classification with low photon counts and shorter measurement times. The VAE is trained over multiple datasets, each with different average photon counts, resulting in a well-behaved latent space that can be used for classification using another neural network (\textit{a classifier}). This approach is robust and maintains classification accuracy despite experimental losses, such as finite collection efficiency and non-unity quantum efficiency, and finite number of the detectors. Moreover, the transfer learning capabilities of VAE allow the successful classification of data of any quality with a \textit{single} trained model. Our method has the potential to significantly enhance the classification of quantum light and light sources, even under conditions of poor detection quality. The importance of efficient state classification in quantum technology is increasing, and the methodology introduced, exemplified by SPACs, SPATs, Coherent, Thermal, and mixed states, could pave the way for efficient classification of light sources. Studies~\cite{haroche2006exploring,giovannetti2004quantum} highlight the role of quantum state discrimination in quantum optics experiments and the practical implications of quantum state classification in enhancing measurement precision.

\section{Method} 

Figure 1 shows the simple schematic of photodetection using four click-counting detectors. Here the incoming low flux light (SPACS/SPATS/thermal/coherent/mixed) is redistributed equally among the detectors using 50/50 beamsplitters.
\noindent \subsection{Dataset generation.} We will begin by considering light in pure SPACS and SPATS. The probability of finding $n$ photons in SPACS and SPATS is given by Equations (1,2), respectively~\cite{barnett2018statistics}.

\begin{equation}
P^{(1+)}_{|\alpha\rangle}(n) = \frac{e^{-|\alpha|^{2}}}{1+|\alpha|^{2}}\left[\frac{|\alpha|^{2(n-1)}}{(n-1)!}+|\alpha|^{2}\frac{|\alpha|^{2(n-2)}}{(n-2)!}\right]
\end{equation}
\begin{equation}
P^{(1+)}_{\text{th}}(n) = \frac{\bar{n}^{n-1}}{(1+\bar{n})^{n+1}}\begin{pmatrix}
n \\
1 
\end{pmatrix} (n\geq 1)
\end{equation}
\noindent where (1+) denotes single photon addition to the input coherent and thermal states with initial mean photon number $\bar{n}$. Equation (1) shows a combination of two shifted and scaled Poissonian distributions, where the first and second terms correspond to shifts by one and two photons, respectively, to the initial coherent state with $\bar{n} =|\alpha|^{2}$. On the other hand, Equation (2) corresponds to a negative binomial distribution for the photon number in SPATS, where for $n<1$ the probability is 0. In experiments, we can also generate mixed state between coherent/SPACS and thermal/SPATS. Employing the density operator approach to denote a mixed state, for instance, a mixed state between coherent ($|\alpha\rangle$) and SPACS ($|\alpha,1\rangle$) with the probability $r$ and $(1-r)$ respectively is given by the relation\cite{Agarwal_2012}:
\begin{equation}
    \hat{\rho}_{(|\alpha\rangle;|\alpha,1\rangle)} = r|\alpha\rangle\langle \alpha|+(1-r)|\alpha,1\rangle\langle 1,\alpha|
\end{equation}
Now, the probability of finding $n$ photons in the field is given by
\begin{equation}
    p(n) = \langle \hat{n}|\hat{\rho}_{(|\alpha\rangle;|\alpha,1\rangle)}|\hat{n}\rangle
\end{equation}
where $\hat{n}$ is the number operator. Similar expression can be written for the mixed state between thermal and SPATS. Using Eqs.(1-4), a substantial dataset containing $10^8$ data points was generated, with each bin having a size of $\tau_{b}$. The dataset included bins with statistical photon counts ranging from 0 to 6 for each bin. We chose to limit the upper bound of photon numbers to 6 based on the observation that higher photon probabilities have a minimal effect on classification, especially for small $\bar{n}$. To achieve accuracy, probabilities were computed up to $P(6)$ during the simulation, however, probabilities up to $P(4)$ were sufficient.

\noindent\subsection{Losses in Dataset.} The overall detection efficiency of the system is determined by the combined effect of linear propagation losses and the photodetector quantum efficiency. This can be represented mathematically using Equation (5), where $\mathcal{P}_{r}(n)$ is the probability of detecting $n$ photons due to losses, and $\mathcal{P}_{i}(m)$ is the probability of finding $m$ photons~\cite{alleaume2004photon} in the ideal lossless scenario. The parameter $\eta$ represents the detection efficiency, and the summation accounts for all possible numbers of photons that can be detected.

\begin{equation}
    \mathcal{P}_{r}(n) = \sum _{m = n}^{\infty} \binom{m}{n} \eta ^n (1 - \eta)^{m - n}\mathcal{P}_{i}(m)
\end{equation}

\noindent However, in reality, there exists a deadtime ($\tau_{D}\simeq$100 ns) for typical Si-based single-photon detectors, which means that each detection channel cannot detect more than one photon within this time interval. As a result, a non-linear relationship exists between the detected photon statistics and the source photon statistics, and the number of detected photons in one bin size cannot exceed the number of detectors. To address this issue,  a low photodetection rate was considered ($R \ll \tau_{D}^{-1}$) to reduce the impact of the detector deadtime. By doing so, one can obtain more accurate and reliable data from the detectors.

\noindent \subsection{Simulating $\mathcal{N}$ detectors.} We employed a multinomial expansion approach to account for losses in a system with $\mathcal{N}$ detectors and $m$ photons ~\cite{alleaume2004photon}. In this approach, each sensor $d_i$ was modeled as having a certain number of photons $k_i$ hitting it, where $k_i$ represents the number of photons received by $i${th} detector. Equation (6) describes the multinomial expansion as follows:
\begin{equation}
    (d_1 + d_2 + ... + d_\mathcal{N})^m = \sum_{\sum{k_i} = m \,\& \,k_i\ge 0} \binom{\mathcal{N}}{k_1, k_2, ...,k_\mathcal{N}}\prod_{i=1}^{\mathcal{N}}d_{i}^{k_i}
\end{equation}

\noindent To determine the probability of exactly $n$ photons being detected, the sum of the coefficients of all terms in the multinomial expansion is calculated with exactly $n$ non-zero $k$ and divided by the sum of all coefficients. To generate simulated data, we computed the theoretical probability distribution that accounted for both types of losses. We calculated the theoretical probability of having $n$ photons $P_{i}(n)$ for $0\leq n\leq 20$, and ignored $P_{i}(n)$ for $n>20$ as it's below a certain threshold that can be ignored for the simulation. This is done by ensuring that the  $\sum\limits_{n>20}P_{i}(n) \leq10^{-6}$. We then incorporated the quantum efficiency using Equation (5) and the deadtime loss using Equation (6). The probability of detecting $n$ photons using $\mathcal{N}$ sensors was calculated using the following equations:
\begin{equation}
    S = \{K: k_i > 0 \}
\end{equation}
\begin{equation}
    C(n, j) = \frac{\sum\limits_{\substack{\text{len}(S)=n \ \& \sum k_i=j}} \binom{\mathcal{N}}{k_1, k_2, ..., k_\mathcal{N}}}{\mathcal{N}^j}
\end{equation}
\begin{equation}
    P_{o}(n) = \sum_{j \geq n} C(n, j)*P_{i}(j)
\end{equation}
In the above equations, $S$ represents the set of all combinations of detectors that detect photons. $C(n, j)$ represents the sum of the coefficients of all terms in the multinomial expansion with exactly $n$ non-zero $k$ and with a sum of $j$. In other words, $C(n, j)$ represents if the emitter emitted $j$ photons then the probability of it being received at $n$ detectors (a detector can receive multiple photons, but it cannot state how many. Therefore, we count each detection as a single photon, even if it may have been multiple photons that were received). Finally, $P_{o}(n)$ represents the probability of detecting $n$ photons. After calculating the final probability incorporating losses, we generated $10^8$ datapoints based on the probability distribution. We then binned the dataset and calculated $P_{o}(0), P_{o}(1),...,P_{o}(6),$ and $\bar{n}_{o}$, which represent the probabilities of detecting 0 to 6 photons and the mean number of photons observed, respectively. These values, along with the bin data for each bin, were used as training data. A total of 2000 training data points for each bin size.

\noindent\subsection{Network and Training.} The VAE is an unsupervised deep learning technique used for generating new data from an existing dataset by extracting a compact and organized representation of the input data \cite{kingma2013auto}. The VAE model has two main components: \textit{encoder} and  \textit{decoder}. The encoder takes input data and generates a latent representation that represents the input data in a condensed and structured format. On the other hand, the decoder takes the latent representation as input and reconstructs the original input, aiming to produce an accurate reconstruction of the input data. The VAE model employed in this study, as shown in Figure 2 (a), is composed of three primary components: the \textit{encoder}, the \textit{decoder}, and a \textit{classifier}. To begin, the \textit{encoder} receives the photon number probabilities represented by P(X) at the input nodes, where X=(0,1,...,4). The input data is then processed through a series of dense layers with a scaled exponential linear unit (SELU) activation function, batch normalization, and dropout, collectively forming the \textit{encoder} network. The final dense layer output comprises of the mean $\mu$ and $\sigma$ values of the latent representation. These values are utilized to sample from a normal distribution, which is subsequently used as the latent representation (\textit{Z}), also known as the bottleneck of the VAE.

The latent representation (\textit{Z}), which is obtained from the bottleneck layer, is fed into the decoder to reconstruct the input data. The decoder is made up of dense layers with SELU activation function, batch normalization, anwrited dropout, similar to the encoder. The final dense layer produces the reconstructed input data. During training, both the encoder and decoder networks are trained to minimize the reconstruction loss. After obtaining the latent representation, the \textit{classifier} takes it as input and predicts the class label. In this study, the classifier is trained to classify SPATS or SPACS by minimizing the classification loss. It uses the latent representation $Z$ to produce a scalar output that indicates whether the input data belongs to a certain class or not. The classifier comprises dense layers with leaky rectified linear unit (LeakyReLU) activation function, batch normalization, and dropout. The VAE model used in this study employs a loss function that combines two terms: reconstruction loss and Kullback-Leibler (KL) divergence loss. The reconstruction loss assesses the dissimilarity between the original input data and the reconstructed output from the decoder. On the other hand, the KL divergence loss measures the variation between the distribution of the latent representation generated by the encoder and a standard normal distribution. By using this loss function, the VAE model is trained to produce diverse and realistic data by encouraging the encoder to produce latent representations that are close to a standard normal distribution. Additionally, the sharing of the bottleneck layer between the VAE and the classification model (\textit{classifier}) allows the VAE to learn a latent representation that is appropriate for both reconstruction and classification tasks. This is particularly useful for complex data that have high dimensions or structures. During the training process, both the VAE and the classification model losses are simultaneously minimized by summing the VAE loss and the classification loss.
\begin{equation}
L_{total} = L_{recon} + L_{KL} + L_{BCE}
\end{equation}
where $L_{recon}$ is the reconstruction loss, $L_{KL}$ is the KL divergence loss, and $L_{BCE}$ is the binary cross entropy loss for the classification task. 

\noindent Here, reconstruction loss~\cite{kingma2013auto}:
\begin{equation}
L_{\mathrm{recon}} = -\frac{1}{N} \frac{1}{d}\sum_{i=1}^N \sum_{j=1}^d (x_{ij} - \hat{x}_{ij})^2,
\end{equation}
where $N$ is the number of samples, $d$ is the dimensionality of the data, $x_{ij}$ is the $j$-th component of the $i$-th input sample, and $\hat{x}_{ij}$ is the corresponding reconstructed value.

\noindent KL divergence loss~\cite{kingma2013auto}:
\begin{equation}
L_{\mathrm{KL}} = -\frac{1}{2N}\sum_{i=1}^N \sum_{j=1}^d [1 + \log(\sigma_j^2) - \mu_j^2 - \sigma_j^2],
\end{equation}
where $\mu_j$ and $\sigma_j^2$ are the mean and variance of the $j$-th component of the latent representation, respectively. 

\noindent Binary cross entropy loss~\cite{goodfellow2016deep}:
\begin{equation}
L_{\mathrm{BCE}} = -\frac{1}{N_{\mathrm{L}}}\sum_{i=1}^{N_{\mathrm{L}}} \left[y_i \log(\hat{y}_i) + (1-y_i)\log(1-\hat{y}_i)\right],
\end{equation}
where $N_{\mathrm{L}}$ is the number of labeled samples, $y_i$ is the true class label of the $i$-th labeled sample (either 0 or 1), and $\hat{y}_i$ is the predicted probability of the $i$-th sample belonging to class 1.

\noindent In this VAE model, various hyperparameters were utilized to achieve the desired performance. These hyperparameters include the number of nodes in the dense layers, the dropout rate, the type of activation functions, and the learning rate. The encoder consists of dense layers with 16, 32, 64, 32, and 16 nodes, respectively, while the decoder has 8, 16, 32, 16, and input nodes. A dropout rate of 0.2 was implemented in the encoder, decoder, and classifier. The encoder and decoder were both set to use the SELU activation function, while the classifier used the LeakyReLU activation function. During training, a batch size of 512 was used with a learning rate of 0.001 for the VAE model. Moreover, by employing a fixed seed for the random initialization of the VAE, we ensure consistent initial conditions for each run, thus aiding in replicable results and mitigating the risk of convergence to local minima (see Supplementary Information).
\section{Simulation Experiments}
\noindent \subsection{Lossless Data} Transfer learning is a deep learning training methodology where features learned from one dataset are transferred and applied to another related dataset. Here transfer learning was employed to improve the accuracy and generalization of the VAE model trained on the lossless dataset, particularly when dealing with different bin sizes. Initial training of VAE model is done on a larger bin size of 100 observations per bin. This choice was made to capture more information about the data distribution. During this training phase, the encoder, decoder, and classifier networks were optimized to reconstruct the input data accurately and classify it between SPATS and SPACS. After the initial training on the larger bin-sized data, the model was fine-tuned using smaller bin-sized data. This process involved taking the pre-trained weights of the model obtained from the initial training and continuing the training process using the smaller bin-sized data. By fine-tuning the model on smaller bin sizes, it allows the model to adapt and improve its performance specifically on data with smaller bin sizes. The fine-tuning process enables the model to transfer the knowledge it learned from the initial training on larger bin sizes to the smaller bin-sized data. The pre-trained weights of the model serve as a starting point for the fine-tuning process. As the model continues training on the smaller bin-sized data, it adjusts its parameters to better suit the characteristics and distributions of the smaller bin sizes. This knowledge transfer helps improve the accuracy and performance of the model on smaller bin-sized data. Finally, validation is done on test dataset to evaluate the model's ability to generalize across different bin sizes. This test dataset likely contained smaller bin sizes than the training data. The model's performance on this test dataset was assessed to determine its generalization capability. The model performs well on the test dataset with smaller bin sizes, indicating the effectiveness of the transfer learning approach in improving the model's accuracy and robustness across different bin sizes. By combining the initial training on larger bin sizes, fine-tuning with smaller bin-sized data, and validating the model's performance on a test dataset, this transfer learning approach helped enhance the accuracy and generalization of the VAE model. The pseudo-code implementation is mentioned in Algorithm 1.

\begin{algorithm}[h]
\caption{{\bf Lossless data, Ideal case} \label{Algorithm1}}
\vspace{2mm}
\textit{Training Phase}: \\
\KwIn{Probability distribution $X: $ $[P(0), P(1), P(2), P(3), P(4)]$}
\KwOut{Class $Y$ and reconstruction of $X$ as $\hat{X}$}
1.  Predicts class and reconstructs the $X$ as $\hat{X}$  \\
2. Compute BCE loss between predicted class and ground truth class (SPACS/SPATS)\\
3. Compute reconstruction loss (MSE) and KL Divergence loss \cite{kingma2013auto}\\
4. Optimize the model using all three losses until convergence.

\textit{Testing Phase}: \\
\KwIn{Probability distribution $X: $ $[P(0), P(1), P(2), P(3), P(4)]$}
\KwOut{Class $Y$ and Embedding of $X$ from bottleneck layer  as $Z$}
\end{algorithm}

\noindent \subsection{Data with losses} The training scheme for the VAE model, which takes into account losses, begins with a data processing step that involves carefully selecting training data. This step is crucial to ensure the applicability of the model to various bin sizes and efficiencies. To improve the generalization capabilities of the model, we have chosen a larger bin size of 200 observations. This decision provides the model with a wider range for classifying data, eliminating the need for multiple models to handle different use cases or minor changes in losses. To enhance the model's flexibility in handling diverse data, we have introduced an additional input parameter called the observed average photon number ($\bar{n}_{obs}$), which is incorporated alongside the probability distribution. By including $\bar{n}_{obs}$ as an input, the model becomes capable of handling a wide range of observable average photon number values, regardless of the corresponding theoretical $\bar{n}_{the}$ and quantum efficiency. This accounts for the effect of quantum loss in the system, as the correlation between $\bar{n}_{obs}$ and  $\bar{n}_{the}$ captures this relationship. In order to generate the training dataset, we determine the range of the target observable average photon number ($\bar{n}_{obs}$). From this range, we select corresponding theoretical $\bar{n}_{the}$ values and quantum efficiencies. By plotting observable $\bar{n}_{obs}$ against theoretical $\bar{n}_{the}$, we can choose the desired theoretical $\bar{n}_{the}$ and quantum efficiency pairs. This approach allows the model to effectively generalize across different observable $\bar{n}_{obs}$ values and account for variations in theoretical $\bar{n}_{the}$ and quantum efficiency. Subsequently, we train the VAE model using this dataset, incorporating the observed $\bar{n}_{obs}$ and the probability distribution. The encoder, decoder, and classifier networks of the model are optimized to minimize the reconstruction loss and classification loss. This training process equips the model with the ability to accurately reconstruct input data while effectively classifying between SPATS and SPACS. For detailed information on the implementation of this training scheme, please refer to Algorithm 2.

\begin{algorithm}[h]
\caption{{\bf Incorporation of observed average photon number,~$\bar{n}_{obs}$ } \label{Algorithm2}}
\vspace{2mm}
\textit{Training Phase}: \\
\KwIn{Probability distribution and $ \overline{n}_{obs} , X: \left[P(0), P(1), P(2), P(3), P(4), \overline{n}_{obs} \right]$}
\KwOut{Class $Y$ and reconstruction of $X$ as $\hat{X}$}
1. Predicts class and reconstructs the $X$ as $\hat{X}$  \\
2. Compute Binary Cross-entropy loss between predicted class and ground truth class (SPACS/SPATS)\\
3. Compute reconstruction loss (MSE) and KL Divergence loss \cite{kingma2013auto}\\
4. Optimize the model using all three losses until convergence.

\textit{Testing Phase}: \\
\KwIn{Probability distribution and $\overline{n}_{obs}, X: \left[P(0), P(1), P(2), P(3), P(4), \overline{n}_{obs} \right]$}
\KwOut{Class $Y$ and Embedding of $X$ from bottleneck layer  as $Z$}
\end{algorithm}

\section{RESULTS AND DISCUSSION}
\noindent Figure \ref{Figures:embedding} presents a comparison between two methods for analyzing the probability distribution of photon statistics of SPATS and SPACS. Initially, we employed the t-Distributed Stochastic Neighbor Embedding (t-SNE) algorithm, a popular multidimensional scaling technique \cite{van2008visualizing}, to visually represent our high-dimensional data. The purpose of t-SNE is to reduce dimensions while preserving the similarity structure within the data, with the expectation that it would allow us to differentiate between the two classes of photon statistics. After running t-SNE for 1000 iterations, we obtained a 3-dimensional representation that revealed a significant overlap between the SPATS and SPACS classes. t-SNE has inherent limitations related to the possibility of different global structures within the data. This variation in global structures could result in the convergence of data points from different classes, leading to overlapping clusters. Furthermore, t-SNE is sensitive to its hyperparameters, and the selection of these parameters can greatly affect the resulting visualization. The overlap observed in our analysis may, in part, be attributed to suboptimal parameter settings. Adjusting these parameters, such as the perplexity value or the learning rate, yields such best-observed separability between the two classes. Moreover, the nature of our dataset, which encompasses different average photon numbers and measurement bin sizes, poses a challenge for t-SNE. The changing photon statistics information across these variations makes it difficult for t-SNE to effectively capture such features. As a consequence, t-SNE proves to be inefficient in representing the diverse range of photon statistics present in the data. The 2D t-SNE visualization of full photon statistics data, distinguished by different photon counts and bin sizes higlights substantial overlap between SPATS and SPACS classes, complicating effective classification (see Supplementary). In contrast, the implementation of the VAE method offers a more distinctive representation of the dataset within the latent space. The VAE model was trained on multiple datasets, encompassing various average photon numbers and measurement bin sizes. This training approach enables the VAE to capture a broader range of photon statistics, resulting in a latent space with distinguishable features and discernible boundaries. Consequently, the VAE method demonstrates superiority in handling such complex data for classification purposes.

\noindent In Figure \ref{Figures:VAE}(b), we explore the relationship between classification accuracy and bin size using lossless data, with an average photon number ($\bar{n}$) = 1.3. As the bin size increases, so does the accuracy, indicative of an intrinsic relationship between these two variables. However, the accuracy plateaus beyond a certain bin size. This saturation effect can be attributed to the inherent nature of photon statistics; larger bins, corresponding to longer observation times, allow for a more precise representation of the photon distribution. Yet, beyond a certain point, the additional information gained does not improve the ability of the model to classify the sources. Despite this, it is worth highlighting that our classification algorithm performs robustly even with smaller bins. This is demonstrated by the maintaining of over 65$\%$ accuracy using a bin size = 50 with a single model trained across multiple datasets of average photon number and bin size and then tested on $\bar{n}$ = 1.3. This promising result suggests the algorithm's potential efficiency even when dealing with sparse data points, which might be especially relevant in real-world applications where high-volume data collection is not always feasible.

\noindent Figure \ref{Figures:loss}(a) provides insights into the relationship between the theoretical mean photon number $\bar{n}_{the}$ and the corresponding observed $\bar{n}_{obs}$ values for four sensors, each characterized by a specific quantum efficiency. Notably, at higher $\bar{n}$ values, saturation occurs, likely due to the fixed number of sensors utilized in the simulation. This saturation phenomenon emphasizes the significance of carefully selecting the number of sensors in the experimental setup to mitigate the sensor ``dead time'' effect and maximize the extraction of information from the light source. In Figure \ref{Figures:loss}(b), we investigate the accuracy of the model as a function of quantum efficiency, using four sensors and a fixed bin size of 200. Remarkably, the model demonstrates consistent performance across a range of quantum efficiencies. However, a notable degradation in performance occurs when the quantum efficiency falls significantly below 0.2, which is a consequence of using a single photon source. This deterioration suggests the presence of a critical threshold below which the signal-to-noise ratio may decline to a level where distinguishing between the probability distributions for SPACS and SPATS becomes challenging, leading to erroneous classification.

\noindent Next, we present a performance analysis shows performance analysis of the \textit{classifier} network within the VAE framework, demonstrating its accuracy across different bin sizes (Figure \ref{Figures:performance}) and observed mean photon numbers, $\bar{n}_{obs}$ (Figure \ref{Figures:acc}). The VAE model was trained using simulated data generated for a theoretical mean photon number of $\bar{n}_{the}$ = 1.9 and a fixed bin size of 200. Notably, during the data generation step, we incorporated loss effects by considering varying quantum efficiencies of 0.9, 0.8, and 0.6. By incorporating these different quantum efficiencies, we constructed a diverse training dataset that encompassed a broad range of observed mean photon number values, ranging from 1.45 to 1.16. This deliberate variation aimed to enhance the ability of the model to generalize well across different scenarios. Remarkably, our trained model exhibited exceptional generalization capabilities, consistently achieving high classification performance across varying bin sizes and observed mean photon numbers. This robust performance was particularly noteworthy given the relatively limited size of our training dataset. The ability of  model to maintain its classification performance despite changes in bin size and observed mean photon numbers implies its adaptability via transfer learning which is a crucial feature for practical implementation, as real-world conditions often exhibit inherent variations. In Figure \ref{Figures:mix}, we demonstrate the architecture's adaptability to new tasks. Here, we analyze four different types of photon sources: mix-SPAC (a mixture of coherent and SPAC sources, where the Y-axis represents its mix ratio), mix-SPAT (a mixture of thermal and SPAT sources, where the X-axis represents its mix ratio), Coherent, and Thermal. As the mix ratio approaches 1, mix-SPAC becomes predominantly Coherent, and mix-SPAT becomes predominantly Thermal, as indicated by Eqs. (3,4). This results in low classification accuracy, as expected. However, the architecture achieves over 90$\%$ for mix ratios less than 0.8, demonstrating its robustness in the classification task.

\section{Conclusions}

\noindent The development of deep learning-based algorithms has demonstrated great potential in the field of quantum optics. Here, a deep learning-based method was proposed and implemented for the efficient classification of non-classical states of light, specifically SPACS and SPATS, which are valuable for various applications such as high-bit-rate communication, and quantum state engineering. The proposed method employed a semi-supervised learning approach using a VAE to map photon statistics features to a lower-dimensional latent space. The VAE was trained over multiple datasets with different average photon counts, resulting in a well-behaved latent space that was robust and maintained classification accuracy even under conditions of poor detection quality. The simulation experimental results showed that the VAE was successful in distinctly representing the dataset in the latent space, and the classification algorithm performed well even with very few observations through transfer learning approach. The proposed method has the potential to significantly enhance the classification of quantum light sources~\cite{akbari2021temperature,jha2021nanoscale,akbari2022lifetime}, especially under conditions of poor detection quality, which is an important step toward the development of practical quantum communication, sensing, and computing technologies. 

\section*{acknowledgements}
\noindent A.K. acknowledges funding support from the Department of Science and Technology via the grants: CRG/2022/001170, ECR/2018/001485, and DST/NM/NS-2018/49. P. K. Jha acknowledges the unrestricted gift from Google and the Syracuse University Start-up Funds, which partially supported this project.

\section*{Author Information}
\noindent $^\ast$To whom all correspondence should be addressed. pkjha@syr.edu\\
\noindent $^\dagger$M. Bhupati and A. Mall contributed equally to this work

\newpage
\bibliography{reference}

\newpage

\begin{figure}[h]
\centering
\includegraphics[scale=0.68]{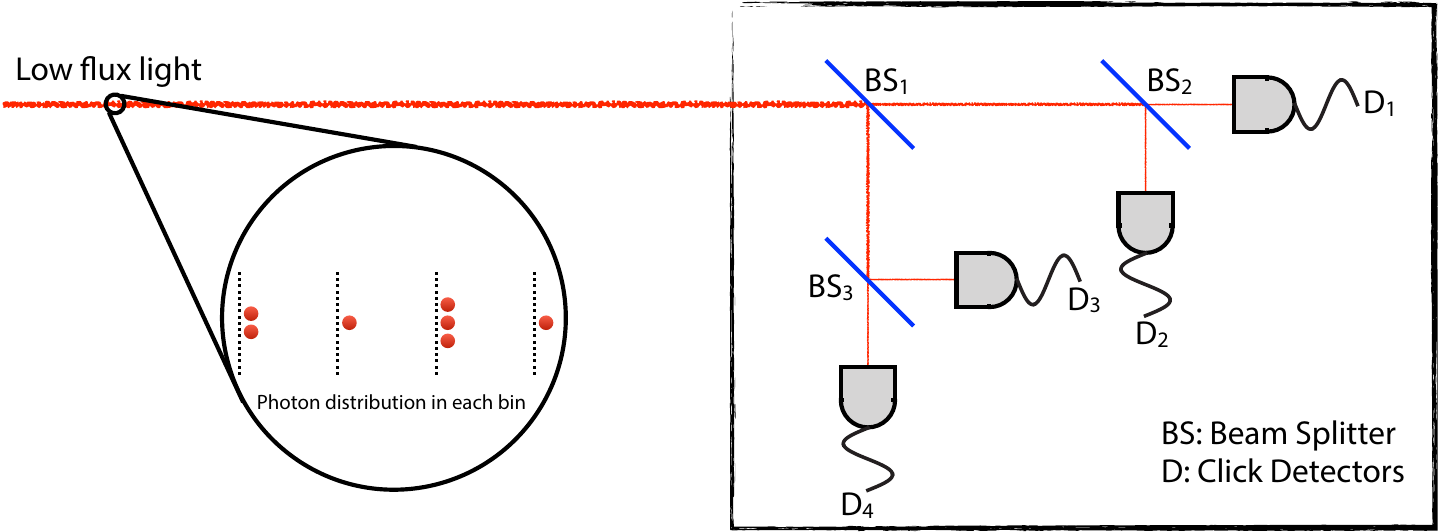}
\caption{Experimental setup illustrating the division of a single incoming light beam into four equal paths using three beamsplitters. Each path directs light toward a dedicated click detector, registering the number of photons detected for a single instance. This process is repeated for \textit{n} trials, where \textit{n} represents the bin size, yielding a dataset of click counts. Analysis of this data produces a probability distribution showcasing the likelihood of detecting a specific number of photons across all \textit{n} samples.}
\label{Figures:schematic}
\end{figure}

\newpage
\begin{figure}[h]
\centering
\includegraphics[scale=1]{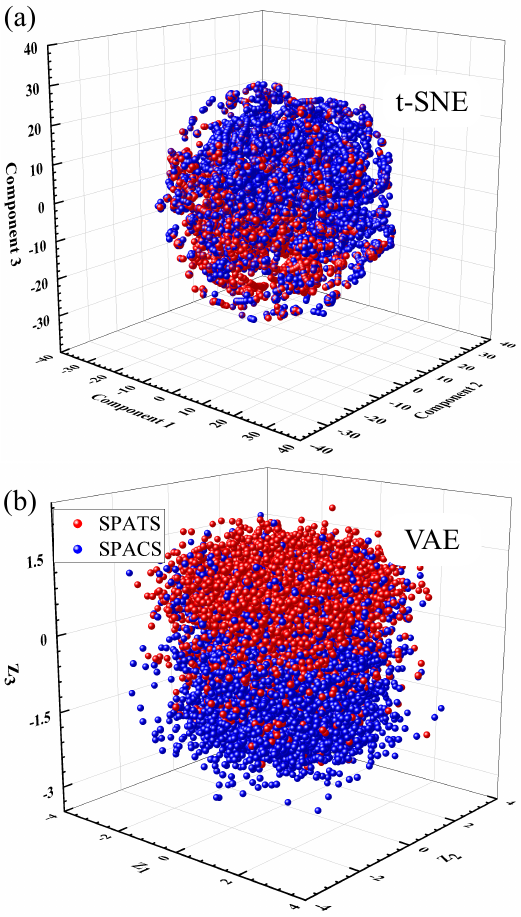}
\caption{Distinct representation of quantum states of light in low-dimensional embedding space. The low-dimensional embedding spaces for photon statistics data, showcase distinct representations of for SPATS $\&$ SPACS across multiple average photon numbers. (a) The t-SNE model's 3-component representation illustrates a dense overlapping of data points clusters. (b) The 3D-latent space of the VAE model reveals two distinct data clusters with good separation.}
\label{Figures:embedding}
\end{figure}


\newpage
\begin{figure}[h]
\centering
\includegraphics[scale=1]{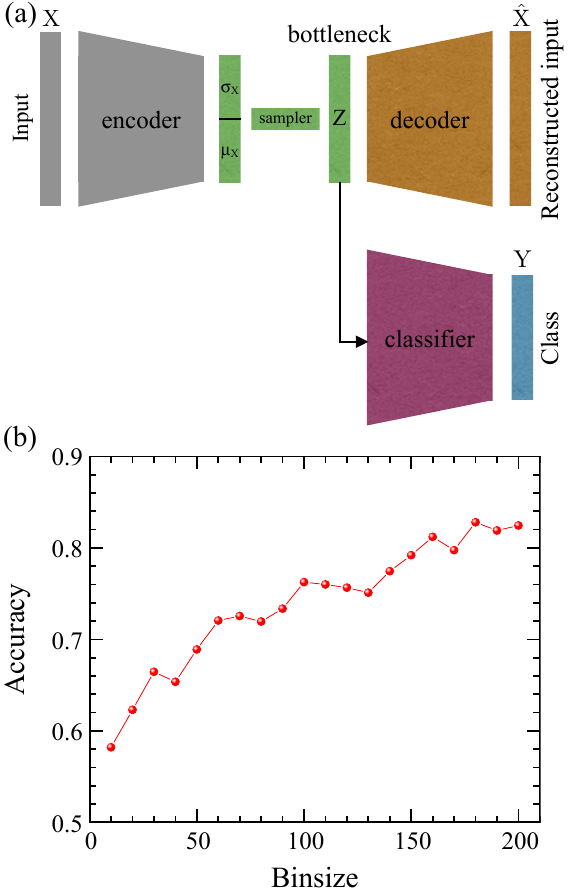}
\caption{Architecture and accuracy of introduced VAE. (a) Schematic representation of the VAE combined with a \textit{classifier} network. The VAE consists of three main components: an encoder, a decoder, and a classifier. The encoder takes the input data, denoted as $X$ which comprises probabilities $P_n$ of having $n$ photons in observation. Here, $X$ is defined as $[P_0, P_1, P_2, \ldots, P_n]$, where $n$ represents the number of sensors, and generates a latent representation. The decoder takes this latent representation as input and reconstructs the original data, resulting in $\hat{X}$. The classifier utilizes the latent representation $Z$ obtained from the bottleneck layer and predicts the class label for SPATS $\&$ SPACS. (b) Investigation of the impact of bin size on the classification accuracy of the \textit{classifier} in the VAE model. The classification accuracy is assessed for varying numbers of binsizes, considering a simulated dataset with an average photon number ($\bar{n}$) = 1.3, without any loss effects.}
\label{Figures:VAE}
\end{figure}

\newpage
\begin{figure}[h]
\centering
\includegraphics[scale=1]{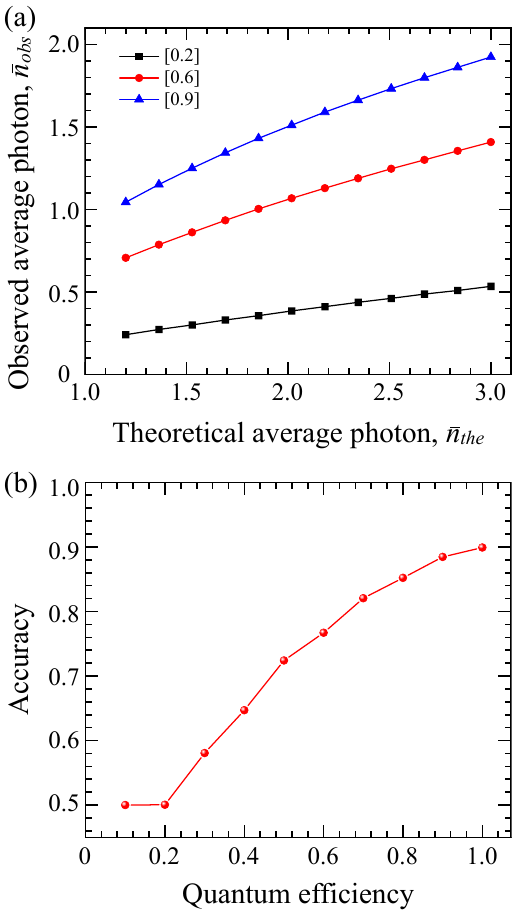}
\caption{Quantum loss and dead time effects on classification accuracy. (a) Illustrates the relationship between the observed average photon number ($\bar{n}_{obs}$) and the theoretical average photon number ($\bar{n}_{the}$) for a fixed quantum efficiency of four sensors. In the simulations, an incoming beam was split into two equal beams using a beam splitter, each having an equal probability of containing a photon. Subsequently, another layer of beam splitters divided these two beams into four beams, each with an equal probability of containing a photon. (b) Depicts the accuracy of the \textit{classifier} in the VAE model as a function of quantum efficiency for a ${\bar{n}_{obs}}$ = 1.9. As the observed average photon number increases with higher quantum efficiency, the \textit{classifier} achieves improved accuracy.}
\label{Figures:loss}
\end{figure}

\newpage
\begin{figure}[h]
\centering
\includegraphics[scale=1]{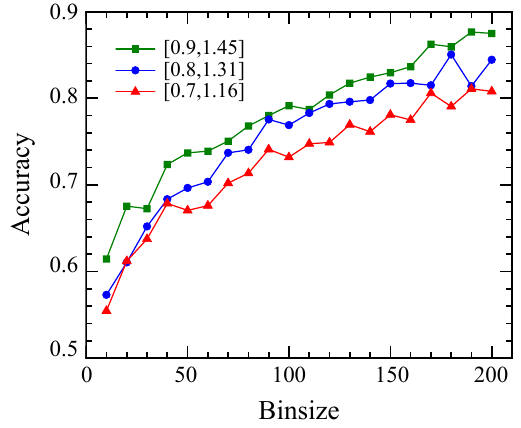}
\caption{Generalization of VAE performance across bin sizes and average observed photon number. The plot shows the relationship between the accuracy of the classifier in the VAE model and the number of binsizes in the simulated experiment. The curves depict the performance at an average theoretical photon number ($\bar{n}_{the}$) of 1.9, where different quantum efficiencies yield varying observed average photon numbers ($\bar{n}_{obs}$). Each curve's color represents the accuracy for different combinations of quantum efficiency and $\bar{n}_{obs}$.}
\label{Figures:performance}
\end{figure}

\newpage
\begin{figure}[h]
\centering
\includegraphics[scale=1.0]{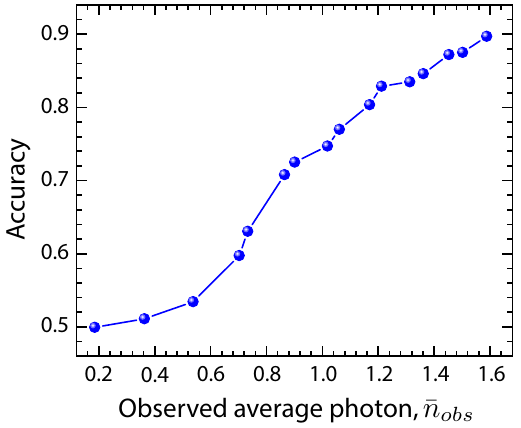}

\caption{The accuracy is directly dependent on $\bar{n}_{obs}$, as we are dealing with a single photon-added source. When $\bar{n} < 1$, it becomes very difficult to distinguish, as it is mostly noise and the loss is too significant to identify the source. As $\bar{n}_{obs}$ increases, the classification accuracy saturates around 90$\%$. (binsize = 200, $\bar{n}_{the} > 1$ (single photon added source), different quantum efficiencies for different $\bar{n}_{obs}$)}
\label{Figures:acc}
\end{figure}

\newpage
\begin{figure}[h]
\centering
\includegraphics[scale=1.5]{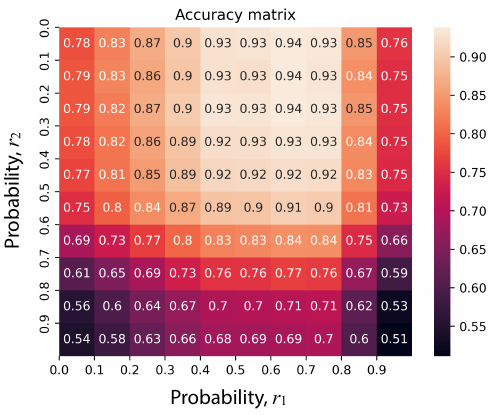}
\caption{The plot of classification accuracy among all four classes: mixSPAC, mixSPAT, Coherent, and Thermal for given values of $r_1$ and $r_2$. Here the x-axis represents mixSPAT (a mixture of Thermal and SPAT with the probability of $r_1$), and the y-axis represents mixSPAC (a mixture of Coherent and SPAC with the probability of $r_2$). The strength of the proposed architecture lies in its adaptability to new data. For this experiment, the observed mean photon number ($\bar{n}$) was fixed at 1.3, and the bin size was set to 200. We observe that the model can adapt to new classes while maintaining similar performance for SPAT and SPAC with $\bar{n}_{obs} = 1.3$ as shown in Figure \ref{Figures:acc}.}
\label{Figures:mix}
\end{figure}

\end{document}